\documentclass[final]{svjour3}
\usepackage{graphicx}
\usepackage{rotating}
\usepackage{amssymb}
\usepackage{amsmath}
\usepackage{mathptmx}
\usepackage[numbers]{natbib}
\usepackage{bm}

\makeatletter
\journalname{Journal of Low Temperature Physics}

\def\lesssim{\ \raise.3ex\hbox{$<$}\kern-0.8em\lower.7ex\hbox{$\sim$}\ }
\def\gesim{\ \raise.3ex\hbox{$>$}\kern-0.8em\lower.7ex\hbox{$\sim$}\ }
\bibpunct{}{}{,}{s}{}{,}
\begin{document}
\title{Rashbon bound states associated with a spherical spin-orbit coupling in an ultracold Fermi gas with an $s$-wave interaction}
\titlerunning{Rashbons in a spin-orbit coupled Fermi gas}
\author{T. Yamaguchi  \and D. Inotani \and Y. Ohashi}
\institute{Department of Physics, Faculty of Science and Technology, Keio University,\\ 3-14-1, Hiyoshi, Kohoku-ku, Yokohama 223-8522, Japan\\
Tel.:+81-45-566-1454 Fax:+81-45-566-1672\\
\email{tmatuura@rk.phys.keio.ac.jp}}
\date{\today}
\maketitle
\begin{abstract}
We investigate the formation of rashbon bound states and strong-coupling effects in an ultracold Fermi gas with a spherical spin-orbit interaction, $H_{\rm so}=\lambda{\bm p}\cdot{\bf \sigma}$ (where ${\bf \sigma}=(\sigma_x,\sigma_y,\sigma_z)$ are Pauli matrices). Extending the strong-coupling theory developed by Nozi\`eres and Schmitt-Rink (NSR) to include this spin-orbit coupling, we determine the superfluid phase transition temperature $T_{\rm c}$, as functions of the strength of a pairing interaction $U_s$, as well as the spin-orbit coupling strength $\lambda$. Evaluating poles of the NSR particle-particle scattering matrix describing fluctuations in the Cooper channel, we clarify the region where rashbon bound states dominate the superfluid phase transition in the $U_{s}$-$\lambda$ phase diagram. Since the antisymmetric spin-orbit interaction $H_{\rm so}$ breaks the inversion symmetry of the system, rashbon bound states naturally have, not only a spin-singlet and even-parity symmetry, but also a spin-triplet and odd-parity symmetry. Thus, our results would be also useful for the study of this parity mixing effect in the BCS-BEC crossover regime of a spin-orbit coupled Fermi gas.
\par
\noindent
\end{abstract}
\par
\section{Introduction}
\par
Recently, an ultracold Fermi gas has attracted much attention as a useful quantum simulator to study various many-body problems in strongly correlated Fermi systems\cite{cross1,cross2,cross3,cross4}. Using a tunable $s$-wave pairing interacting associated with a Feshbach resonance, we can now study the $s$-wave Fermi superfluid state from the weak-interacting BCS (Bardeen-Cooper-Schrieffer) regime to the strong-coupling BEC (Bose-Einstein condensation) limit in a unified manner\cite{swave1,swave2,swave3,swave4,swave5,swave6,swave7,swave8,swave9,swave10,swave11}. In addition, the recent experimental progress on artificial gauge field technique\cite{soc1,soc2,soc3,soc4} has enabled us to introduce a spin-orbit interaction to a Fermi gas, in spite of the fact that this is a neutral gas. In particular, since an antisymmetric spin-orbit interaction breaks the spatial inversion symmetry, the so-called parity-mixing effect has been discussed\cite{2b1,2b2,2b3,2b4,2b5,HuiHu}. As an interesting application of this unique phenomenon, the possibility of a $p$-wave superfluid Fermi gas by using this effect has been proposed\cite{mypaper}. In this regard, we briefly note that, while a tunable $p$-wave pairing interaction associated with a $p$-wave Feshbach resonance has already been realized\cite{pwave1,pwave2,pwave3}, the $p$-wave superfluid state has not been reported yet. Since this unconventional pairing state has been realized in various Fermi systems, such as liquid $^{3}$He\cite{3he1,3he2}, as well as heavy-fermion superconductors\cite{heavy1,heavy2,heavy3}, the realization of a $p$-wave superfluid Fermi gas with a tunable interaction would contribute to the further development of unconventional superfluid physics. 
\par
Besides the parity-mixing effect, it is also known that a spin-orbit interaction produces a two-body bound state even in the weak-coupling BCS regime (where the ordinary contact-type $s$-wave pairing interaction never produces a bound state). This novel bound state is sometimes referred to as the rashbon in the literature, and it has also been pointed out that they enhance the superfluid phase transition temperature $T_{\rm c}$ in the weak-coupling BCS regime, as in the case of the strong-coupling BEC regime (where the superfluid phase transition is dominated by tightly bound two-body states). Thus, when we consider the phase diagram of a spin-orbit coupled Fermi gas with respect to the strength of a $s$-wave pairing interaction and the strength of a spin-orbit interaction, it is an interesting problem to identify the region where the superfluid phase transition can be well described by the BEC of rashbons. 
\par
In this paper, we investigate a spin-orbit coupled ultracold Fermi gas in the BCS-BEC crossover region at $T_{\rm c}$. To simplify our discussions, we treat the case of a spherical spin-orbit coupling, $H_{\rm so}=\lambda{\bm p}\cdot{\bf \sigma}$ (where ${\bf \sigma}=(\sigma_x,\sigma_y,\sigma_z)$ are Pauli matrices). Within the framework of the BCS-BEC crossover theory developed by Nozi\`eres and Schmitt-Rink (NSR)\cite{NSR}, we determine $T_{\rm c}$, and an effective Fermi chemical potential ${\tilde \mu}$ in a consistent manner. Noting that the region with a negative Fermi chemical potential at $T_{\rm c}$ may be regarded as the BEC regime, we divide the phase diagram of this system into the BCS regime and (molecular) BEC regime. In addition, using that the rashbon mass is heavier than the mass of a Cooper pair, we further divide the BEC regime into the ``rashbon-BEC regime" and the ``ordinary BEC regime" where the superfluid phase transition is simply dominated by tightly bound Cooper pairs. Although there is actually no phase transition between, for example, BCS regime and rashbon-BEC regime, this phase diagram would be helpful to grasp the overall behavior of a spin-orbit coupled Fermi gas in the BCS-BEC crossover region. Throughout this paper, we set $\hbar=k_{\rm B}=1$, and the system volume $V$ is taken to be unity, for simplicity.
\par
\par
\section{Formulation}
\par
We consider a spin-orbit coupled two-component Fermi gas with an $s$-wave pairing interaction $-U_s$ ($<0$). In the functional integral formalism\cite{SadeMelo,Negele}, this interacting Fermi system is described by the action,
\begin{equation}
S= \int dx
\Bigl[
{\bar \Psi}(x)
\Bigl[
{\partial \over \partial\tau}
+
{{\hat {\bm p}}^2 \over 2m}-\mu
+
\lambda{\hat {\bm p}}\cdot{\bm \sigma}
\Bigr]
\Psi(x)
-U_{s}
{\bar \psi}_\uparrow(x){\bar \psi}_\downarrow(x)
\psi_\downarrow(x)\psi_\uparrow(x)
\Bigr].
\label{eq.1} 
\end{equation}
Here, we have employed the simplified notations, $x=({\bm r},\tau)$ and $\int dx=\int_0^\beta d\tau\int d{\bm r}$, where $\beta=1/T$ is the inverse temperature. $\psi_\sigma(x)$ and ${\bar \psi}(x)$ is the Grassmann variable and its conjugate, describing a Fermi atom with pseudospin $\sigma=\uparrow,\downarrow$ (denoting two atomic hyperfine states) and with atomic mass $m$. In Eq. (\ref{eq.1}), $\Psi(x)=(\psi_\uparrow,{\bar \psi}_\downarrow)^T$, ${\hat {\bm p}}=-i\nabla$, and $\mu$ is the Fermi chemical potential. $\lambda{\hat {\bm p}}\cdot{\bm \sigma}$ is a spherical spin-orbit interaction with the coupling constant $\lambda$, where ${\bm \sigma}=(\sigma_x,\sigma_y,\sigma_z)$ are Pauli matrices. We briefly note that $\lambda$ can be taken to be positive, without loss of generality. 
\par
For the action $S$ in Eq. (\ref{eq.1}), we evaluate the partition function,
\begin{equation}
Z=\int\prod_{\sigma}\mathcal{D}
{\bar \psi}_\sigma\mathcal{D}\psi_\sigma e^{-S}.
\label{eq.2}
\end{equation}
For this purpose, we conveniently employ the Hubbard-Stratonovich transformation\cite{SadeMelo}, to carry out the functional integrals with respect to $\psi_\sigma(x)$ and ${\bar \psi}_\sigma(x)$, which gives
\begin{equation}
Z=\int \mathcal{D}\Delta^*\mathcal{D}\Delta e^{-S_{\rm eff}(\Delta,\Delta^*)},
\label{eq.3}
\end{equation}
where $\Delta(x)$ is a Cooper-pair field, and the effective action $S_{\rm eff}$ has the form,
\begin{equation}
S_{\rm eff}(\Delta,\Delta^*)= \int dx
{|\Delta(x)|^2 \over U_{s}}-{1 \over 2}\rm{Tr}\ln[-{\hat G}^{-1}].
\label{eq.4}
\end{equation}
Here, 
\begin{equation}
{\hat G}^{-1}(x,x')=
\left(
\begin{array}{cc}
\displaystyle
-{\partial \over \partial\tau}-
\Bigl[{{\hat {\bm p}}^2 \over 2m}-\mu\Bigr]-{\hat H}_{\rm so} & 
\displaystyle
i{\hat \sigma}_y\Delta(x) \\
\displaystyle
-i{\hat \sigma}_y\Delta^*(x)  
& 
\displaystyle
-{\partial \over \partial\tau}+
\Bigl[{{\hat {\bm p}}^2 \over 2m}-\mu\Bigr]
-{\hat H}_{\rm so}^*
\end{array}
\right)
\delta(x-x')
\label{eq.5} 
\end{equation}
is the $4\times 4$-matrix single-particle thermal Green's function in real space\cite{maki,shiba}. 
\par
In the NSR theory\cite{NSR}, the superfluid phase transition temperature $T_{\rm c}$ is conveniently determined from the saddle point condition $\delta S_{\rm eff}/\delta\Delta^*(x)=0$\cite{SadeMelo} with $\Delta(x)\to 0$. The resulting $T_{\rm c}$-equation formally has the same form as the ordinary BCS gap equation at $T_{\rm c}$ as,
\begin{equation}
1={U_s \over 2}
\sum_{{\bm p},\alpha=\pm}
{1 \over 2\xi_{\bm p}^\alpha}
\tanh{\xi_{\bm p}^{\alpha} \over 2T}, 
\label{eq.6}
\end{equation}
where $\xi_{\bm p}^\pm =[p\pm m\lambda]^2/(2m)-{\tilde \mu}$ is the single-particle dispersion in the presence of the spherical spin-orbit interaction. ${\tilde \mu}=\mu+m\lambda^2/2$ is the Fermi chemical potential, measured from the bottom of the lower band. As usual\cite{NSR,SadeMelo}, we solve the $T_{\rm c}$-equation, together with the equation for the total number $N=-\partial \Omega_{\rm NSR}/\partial \mu$ of Fermi atoms. Here, the NSR thermodynamic potential $\Omega_{\rm NSR}=-T\ln Z$ is obtained by expanding the action $S_{\rm eff}$ in Eq. (\ref{eq.4}) around $\Delta(x)=0$ to the quadratic order, which is followed by carrying out functional integrals with respect to $\Delta(x)$ and $\Delta^*(x)$. The resulting NSR number equation is given by
\begin{equation}
N = \sum_{{\bm p},\alpha=\pm}f(\xi_{\bm p}^\alpha)-
T{\partial \over \partial \mu}
\sum_{{\bm q},i\nu_n}\ln
\Bigl[
-\Gamma^{-1}({\textrm{\boldmath $q$}},i\nu_n)
\Bigr]e^{i\nu_{n}\delta},
\label{eq.7}
\end{equation}
where $f(x)$ is the Fermi distribution function, $\nu_n$ is the boson Matsubara frequency, and $\delta$ is an infinitely small positive number. In Eq. (\ref{eq.7}), the first term is just the same form as the number of free Fermi atoms, and the second term describes fluctuation corrections. Here,
\begin{equation}
\Gamma({\bm q},i\nu_n) = 
{
{4\pi a_s \over m}
\over
1+{4\pi a_s \over m}
\left[
\Pi({\bm q},i\nu_n)-\sum_{\bm p}{m \over p^2}
\right]
}, 
\label{eq.8}
\end{equation}
is the particle-particle scattering matrix, and
\begin{equation}
\Pi({\bm q},i\nu_n)
={1 \over 4}\sum_{{\bm p},\alpha,\alpha'=\pm}
{1-f(\xi_{{\bm p}+{\bm q}/2}^\alpha)-f(\xi_{-{\bm p}+{\bm q}/2}^{\alpha'})
\over 
\xi_{{\bm p}+{\bm q}/2}^\alpha+\xi_{-{\bm p}+{\bm q}/2}^{\alpha'}-i\nu_n
}
\Bigl[
1+\alpha\alpha'
{
({\bm p}+{\bm q}/2)\cdot({\bm p}-{\bm q}/2)
\over
|{\bm p}+{\bm q}/2||{\bm p}-{\bm q}/2|
}
\Bigr]
\label{eq.9}
\end{equation}
is the lowest-order pair-correlation function, involving effects of spin-orbit coupling\cite{mypaper}. As in the absence of spin-orbit interaction, the pair-correlation function $\Pi({\bm q},i\nu_n)$ in Eq. (\ref{eq.9}) exhibits the ultraviolet divergence. This singularity, however, has been eliminated in Eq. (\ref{eq.8}), by introducing the $s$-wave scattering length $a_s$, given by\cite{inas}
\begin{equation}
{4\pi a_{s} \over m}
=-{U_{s} \over 1-U_{s}\sum_{\bm p}^{p_{\rm c}}
{m \over p^2}},
\label{eq.9b}
\end{equation}
where $p_{\rm c}$ is a momentum cutoff.
\par
\begin{figure}[t]
\begin{center}
\includegraphics[width=0.9\textwidth]{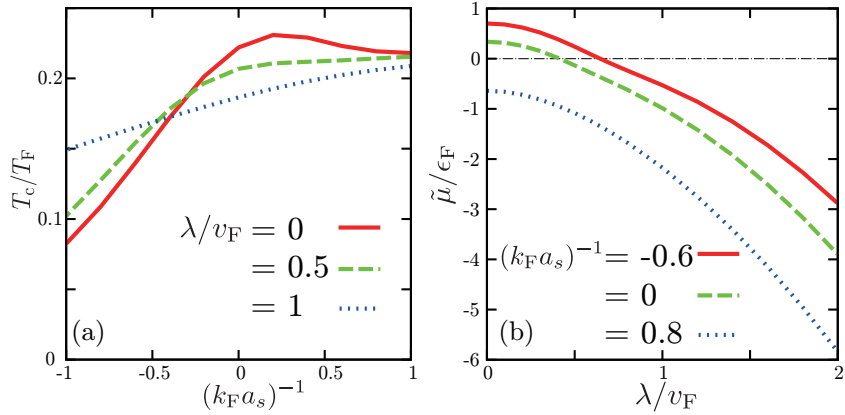}
\caption{ (Color online) (a) Calculated superfluid phase transition temperature $T_{\rm c}$ in a spin-orbit coupled ultracold Fermi gas.  (b) Effective chemical potential ${\tilde \mu}$ at $T_{\rm c}$, as a function of the spin-orbit coupling constant $\lambda$, normalized by the Fermi velocity $v_{\rm F}$.}
\label{fig1}
\end{center}
\end{figure}
\par
\section{Phase diagram of an ultracold Fermi gas with a spherical spin-orbit coupling}
\par
Figure \ref{fig1}(a) shows the superfluid phase transition temperature $T_{\rm c}$ in the BCS-BEC crossover regime of a spin-orbit coupled Fermi gas. In this figure, the spin-orbit interaction is found to enhance $T_{\rm c}$ in the weak-coupling BCS regime ($(k_{\rm F}a_s)^{-1}\lesssim -0.5$)\cite{2b2,mypaper}. As in the strong-coupling BEC regime with $\lambda\sim 0$ (where tightly bound molecules dominate over the superfluid instability), the superfluid phase transition in the weak-coupling BCS regime with a strong spin-orbit interaction is also dominated by bound states (rashbons). Indeed, as shown in Fig. \ref{fig1}(b), the effective Fermi chemical potential ${\tilde \mu}$ gradually deviates from the Fermi energy $\varepsilon_{\rm F}$ with increasing the coupling strength $\lambda$, to be negative in the strong spin-orbit coupling regime, which is just the same as the well-known BCS-BEC crossover behavior of the Fermi chemical potential in the absence of spin-orbit interaction\cite{cross1,cross2,cross3,cross4}. In the latter case, the Fermi chemical potential $\mu$ (Note that ${\tilde \mu}=\mu$ when $\lambda=0$.) approaches half the binding energy $E_{\rm bind}=-1/(ma_s^2)$ of a two-body bound state in the BEC limit\cite{swave11}. Thus, although there is no clear boundary between the BCS-type Fermi superfluid and the BEC of tightly bound molecules, it is convenient to distinguish between the two by the sign of ${\tilde \mu}$.
\par
\begin{figure}
\begin{center}
\includegraphics[width=0.5\textwidth]{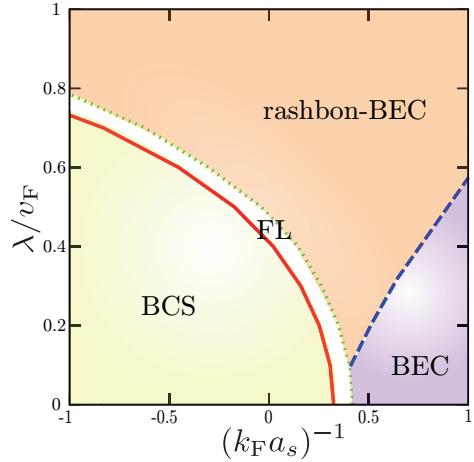}
\caption{(Color online) Boundary between the BCS-type superfluid phase transition (`BCS') and the BEC-type superfluid phase transition (`rashbon-BEC' and `BEC'), determined from vanishing effective Fermi chemical potential, ${\tilde \mu}(T_{\rm c})=0$ (solid line). The dotted line satisfies the condition $T_{\rm c}=2|{\tilde \mu}(T_{\rm c})|$. In the region between the dotted line and the solid line (`FL'), one expects that bound molecules dissociate thermally to some extent. The dashed line is the boundary between the BEC phase transition of the ordinary stable preformed Cooper pairs (`BEC') and that of rashbons (`rashbon-BEC'), which is determined from the condition $M_{\rm{B}}/(2m)=1.02$. We note that the no phase transition occurs at the ``boundaries'' in this figure, but the character of the superfluid phase transition continuously changes across them.}
\label{fig2}
\end{center}
\end{figure}
\par
Using this criterion, we identify the boundary between the region where the BCS-type superfluid phase transition occurs at $T_{\rm c}$ (${\tilde \mu}(T_{\rm c})\ge 0$), and the region where the superfluid phase transition is dominated by bound molecules (${\tilde \mu}(T_{\rm c})<0$), as the solid line in Fig. \ref{fig2}. (The condition ${\tilde \mu}(T_{\rm c})=0$ is satisfied along this line.) When one simply regards $2{\tilde \mu}$ ($<0$) as the binding energy of a bound molecule, we expect that molecules in the latter region (${\tilde \mu}(T_{\rm c}<0)$) are still affected by thermal dissociation when the ``binding energy" $2|{\tilde \mu}(T_{\rm c})|$ is smaller than the thermal energy $T_{\rm c}$\cite{swave10}. Including this effect, we also draw the line which satisfies $T_{\rm c}=2|{\tilde \mu}(T_{\rm c})|$ in Fig. \ref{fig2}. In this case, the region ``FL'' is interpreted as the pairing-fluctuation regime where molecules partially dissociate thermally.
\par
\begin{figure}
\begin{center}
\includegraphics[width=0.9\textwidth]{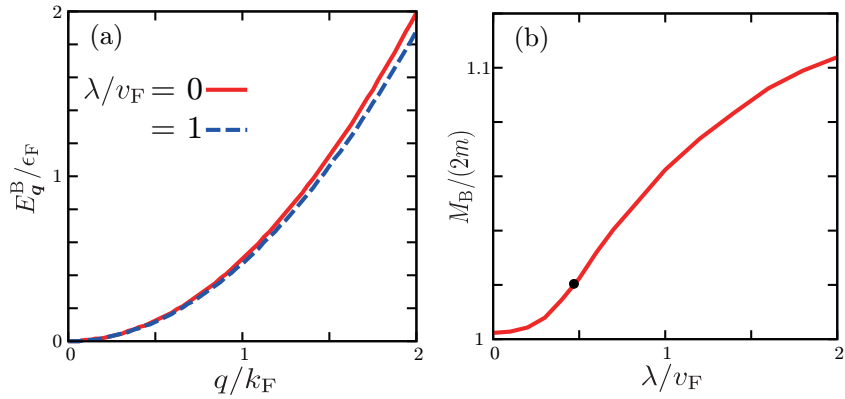}
\caption{ (Color online) (a) Dispersion $E_{\bm q}^{\rm B}$ of a bound molecule at $T_{\rm c}$. (b)  Mass $M_{\rm B}$ of a bound molecule at $T_{\rm c}$ when ${\tilde \mu}<0$. We take $(k_{\rm F}a_s)^{-1}=0.8$. 
The filled circle shows the position at which we determined as the boundary between the ``rashbon-BEC'' and ``BEC'' in Fig. \ref{fig2}. 
}
\label{fig3}
\end{center}
\end{figure}

While the molecular BEC region around $\lambda=0$ is dominated by ordinary preformed Cooper pairs, the molecular region in the weak-coupling regime in Fig. \ref{fig2} is dominated by rashbons. Although there is, of course, no clear difference between these molecular states, it is still convenient to draw a ``boundary" between them based on a physical picture. In this regard, we recall that the mass of a rashbon is known to be different from the mass ($=2m$) of a simple molecule\cite{2b5}. Thus, to use this difference, in this paper, we determine the molecular dispersion ($E_{\bm q}^{\rm B}=q^2/2M_{\rm B}$) at $T_{\rm c}$, by evaluating the pole of the analytic continued particle-particle scattering matrix $\Gamma({\bm q},i\nu_n\to\omega+i\delta)$ in Eq. (\ref{eq.8}). From the molecular dispersion shown in Fig. \ref{fig3}(a), we then evaluate the molecular mass $M_{\rm B}$ shown in Fig. \ref{fig3}(b), to determine the region of ``rashbon-BEC'' in Fig. \ref{fig2}, as the region where $M_{\rm B}/(2m)\ge 1.02$. In Fig. \ref{fig2}, the ``BEC" region is thus dominated by stable preformed Cooper pairs with the molecular mass $M_{\rm B}\simeq 2m$. 
\par
\section{Summary}
\par
To summarize, we have discussed the superfluid phase transition in a spin-orbit coupled $s$-wave ultracold Fermi gas in the BCS-BEC crossover region. In the case of a spherical spin-orbit coupling, from the sign of the effective Fermi chemical potential, we distinguished the region where the system is dominated by stable bound molecule (${\tilde \mu}<0$) from the ordinary BCS state (${\tilde \mu}\ge 0$) in the phase diagram with respect to the interaction strength $-U_s$ and the spin-orbit coupling strength $\lambda$. In addition, using the molecular mass, we also divided the former region into the rashbon-BEC regime and the BEC regime consisting of ordinary preformed Cooper pairs. Although these boundaries involve ambiguity because there is actually no phase transition there, the phase diagram obtained in this paper would be useful in studying how rashbons affect many-body properties of a spin-orbit coupled ultracold Fermi gas.
\par
\begin{acknowledgements}
We thank R. Hanai, H. Tajima, M. Matsumoto and P. van Wyk, for discussions. This work was supported by the KiPAS project in Keio university. H.T. and R.H. were supported by the Japan Society for the Promotion of Science. Y.O. was supported by Grant-in-Aid for Scientific Research from MEXT and JSPS in Japan (No.25400418, No.15H00840).
\end{acknowledgements}


\begin{thebibliography}{99}
\bibitem{cross1} Giorgini, S. Pitaevskii, S. Stringari, Rev. Mod. Phys. \textbf{80}, 1215 (2008). 
\bibitem{cross2} I. Bloch, J. Dalibard, W. Zwerger, Rev. Mod. Phys. \textbf{80}, 885 (2008).
\bibitem{cross3} C. Chin, R. Grimm, P. Julienne, E. Tiesinga, Rev. Mod. Phys. \textbf{82}, 1225 (2010).
\bibitem{cross4} Y. Ohashi, A. Griffin, Phys. Rev. A \textbf{67}, 063612 (2003).
\bibitem{swave1} Y. Ohashi, Phys. Rev. Lett. \textbf{94}, 050403 (2005).
\bibitem{swave2} T.-L. Ho and R. B. Diener, Phys. Rev. Lett. \textbf{94}, 090402 (2005).
\bibitem{swave3} V. Gurarie, L. Radzihovsky, and A. V. Andreev, Phys. Rev. Lett. \textbf{94}, 230403 (2005).
\bibitem{swave4} J. Levinsen, N. R. Cooper, and V. Gurarie, Phys. Rev. Lett. \textbf{99}, 210402 (2007).
\bibitem{swave5} S. S. Botelho and C. A. R. S\'a de Melo, J. Low Temp. Phys. \textbf{140}, 409 (2005).
\bibitem{swave6} M. Iskin and C. A. R. S\'a de Melo, Phys. Rev B \textbf{72}, 224513 (2005).
\bibitem{swave7} E. Grosfeld, N. R. Cooper, A. Stern, and R. Ilan, Phys. Rev. B \textbf{76}, 104516 (2007).
\bibitem{swave8} C.-H. Cheng and S.-K. Yip, Phys. Rev. Lett. \textbf{95}, 070404 (2005).
\bibitem{swave9} R. A. W. Maier, C. Marzok, C. Zimmermann, and Ph. W. Courteille, Phys. Rev. A  \textbf{81}, 064701 (2010).
\bibitem{swave10} R. Watanabe, S. Tsuchiya, and Y. Ohashi, Phys. Rev. A \textbf{82}, 043630 (2010). 
\bibitem{swave11} A. Perali, P. Pieri, G. C. Strinati, and C. Castellani, Phys. Rev. B \textbf{66}, 024510 (2002).
\bibitem{soc1} J. Dalibard, F. Gerbier, G. Juzeliunas, and P. \"Ohberg, Rev. Mod. Phys. \textbf{83}, 1523 (2011).
\bibitem{soc2} Y.-J. Lin, R. L. Compton, A. R. Perry, W. D. Phillips, J. V. Porto, and I. B. Spielman, Phys. Rev. Lett. \textbf{102}, 130401 (2009). 
\bibitem{soc3} L. W. Cheuk, A. T. Sommer, Z. Hadzibabic, T. Yefsah, W. S. Bakr, and M. W. Zwierlein, Phys. Rev. Lett. \textbf{109}, 095302 (2012).
\bibitem{soc4} P. Wang, Z.-Q. Yu, Z. Fu, J. Miao, L. Huang, S. Chai, H. Zhai, and J. Zhang, Phys. Rev. Lett. \textbf{109}, 095301 (2012).
\bibitem{2b1} J. P. Vyasanakere and V. B. Shenoy, Phys. Rev. B \textbf{83}, 094515 (2011), {\it ibid.}, Phys. Rev. A \textbf{86}, 053617 (2012), New J. Phys. \textbf{14}, 043041 (2012).
\bibitem{2b2} Z. Zheng, H. Pu, X. Zou, and G. Guo, Phys. Rev. A \textbf{90}, 063623 (2014).
\bibitem{2b3} J. P. Vyasanakere, S. Zhang, and V. B. Shenoy, Phys. Rev. B \textbf{84}, 014512 (2011).
\bibitem{2b4} Z.-Q. Yu and H. Zhai, Phys. Rev. Lett. \textbf{107}, 195305 (2011).
\bibitem{2b5} L. He and X.-G. Huang, Phys. Rev. B \textbf{86}, 014511 (2012).
\bibitem{HuiHu} L. Jiang, X. Liu, H. Hu, and H. Pu, Phys. Rev. A {\bf 84}, 063618 (2011).
\bibitem{mypaper} T. Yamaguchi, and Y. Ohashi, Phys. Rev. A \textbf{92}, 013615 (2015). 
\bibitem{pwave1} C. A. Regal, C. Ticknor, J. L. Bohn, and D. S. Jin, Phys. Rev. Lett. \textbf{90}, 053201 (2003).
\bibitem{pwave2} C. Ticknor, C. A. Regal, D. S. Jin, and J. L. Bohn, Phys. Rev. A \textbf{69}, 042712 (2004).
\bibitem{pwave3} J. Zhang, E. G. M. van Kempen, T. Bourdel, L. Khaykovich, J. Cubizolles, F. Chevy, M. Teichmann, L. Tarruell, S. J. J. M. F. Kokkelmans, and C. Salomon, Phys. Rev. A \textbf{70}, 030702(R) (2004). 
\bibitem{3he1} A. J. Leggett, Rev. Mod. Phys. 47, 331 (1975).
\bibitem{3he2} D. Vollhardt and P. W\"olfle, {\it The Superfluid Phases of Helium 3}, (Taylor and Francis, New York, 2002).
\bibitem{heavy1} M. Sigrist and K. Ueda, Rev. Mod. Phys. \textbf{63}, 239 (1991).
\bibitem{heavy2} G. R. Stewart, Rev. Mod. Phys. \textbf{56}, 755 (1984).
\bibitem{heavy3} A. P. Mackenzie and Y. Maeno, Rev. Mod. Phys. \textbf{75}, 657 (2003).
\bibitem{NSR} P. Nozi\'eres and S. Schmitt-Rink, J. Low Temp. Phys. \textbf{59}, 195 (1985).
\bibitem{SadeMelo} C. A. R. S\'a de Melo, M. Randeria, and J. R. Engelbrecht, Phys. Rev. Lett. \textbf{71}, 3202 (1993).
\bibitem{Negele} J. W. Negele and H. Orland, {\it Quantum Many-Particle Systems}, (Addison-Wesley, New York,1988) Chapter 2. 
\bibitem{inas} M. Randeria, in {\it Bose-Einstein Condensation}, edited by A. Griffin, D.W. Snoke, and S. Stringari (Cambridge University Press, New York, 1995), p. 355. 
\bibitem{maki} K. Maki, in {\it Superconductivity}, edited by R. D. Parks (Marcel Dekker, New York, 1969) Vol.2, p1035. 
\bibitem{shiba} H. Shiba, Prog. Theor. Phys. \textbf{40}, 435 (1968).
\end{thebibliography}
\end{document}